\documentclass[12pt,psfig,a4]{article}
\usepackage [center,small,bf]{caption}
\usepackage{lipsum}
\usepackage{setspace}
\usepackage{geometry}
\usepackage{tikz}
\usepackage{mathtools}
\usepackage{amsmath}
\usepackage{csquotes}
\usepackage{graphics}
\usepackage{setspace}
\usepackage{parskip}
\newcommand{\Lyx}{L\kern-.1667em\lower.25em\hbox{y}\kern-.125emX\spacefactor1000}

\begin{document}
\bibliographystyle{plain} 
\pagestyle{plain} 
\pagenumbering{arabic}
\date{}
\title{New 2D CA based Image Encryption Scheme and a novel Non-Parametric Test for Pixel Randomness}
\author{ 
   \large BalaSuyambu J \\[-3pt]
    \normalsize Department of Mathematics, IIT Madras, Chennai, Tamil Nadu-600036, India \\[-3pt]
    \normalsize \textcolor{blue} {ma11d002@smail.iitm.ac.in} \\[-3pt]\\
   \large Radha R \\[-3pt]
    \normalsize Department of Mathematics, IIT Madras, Chennai, Tamil Nadu-600036, India  \\[-3pt]
    \normalsize \textcolor{blue} {radharam@iitm.ac.in}\\[-3pt]\\
   \large Rama R \\[-3pt]
    \normalsize Department of Mathematics, IIT Madras, Chennai, Tamil Nadu-600036, India \\[-3pt]
    \normalsize \textcolor{blue} {ramar@iitm.ac.in} \\[-3pt]
}

\maketitle


\begin{abstract} 
In this paper we have proposed a new test for pixel randomness using non-parametric method in statistics. In order to validate this new non-parametric test we have designed an encryption scheme based on 2D cellular automata. The strength of the designed encryption scheme is first assessed by standard methods for security analysis and the pixel randomness is then determined by the newly proposed non-parametric method.

\end{abstract} 


\section{Introduction}

Multimedia images are widely used in internet communications, so the need for securely transmitting these images over networks has become an essential part of the field of data security. Image cryptography plays a vital role in securing confidential images. The purpose of image cryptography is to hide the content of the images by encrypting them so as to make the images unrecognizable to the intruders. One part of this paper deals with a new image cryptosystem using 2D Cellular Automata.\\

	In general, there are two different methods to protect an image; they are (i) image shuffling and (ii) image encryption. Pixels positions are rearranged in image shuffling whereas in image encryption, pixel values and positions are changed. In both the cases it is very essential to check the security of the method. That means the method should be invulnerable to all attacks. Poorly protected images will always provide information about the original image in statistical analysis. If the encrypted image is indistinguishable from the random image, statistical analyses will not have advantage to break. So testing the randomness in the pixels of encrypted image is the state of the art. In the literature already there are different tests for checking randomness for 1D data\cite{ruk}. A number of parametric tests are designed for pixel randomness in shuffled and encrypted images\cite{li,wu,yue}. A non-parametric test is developed in this paper for checking randomness in the image pixels, which is the first of its kind.\\
	John von Neumann proposed a new emerging concept called Cellular Automata (CA)\cite{von}. CA is a discrete model consisting of regular grid of cells, each in one of the finite number of states. According to some fixed rule, the state of each cell will be changed in terms of the state of the current cell and the states of the cells in its neighborhood. Like 1D, higher dimension CA also can be defined\cite{wolf}. It is already proved that some of the rules of CA will be able to generate complex random patterns\cite{wolf1}. In the last two decades 1D and 2D CA’s are used in Cryptography. Lot of research is going on for CA based image cryptography \cite{suya,balas,bala}. The paper is organized as follows: Section 2 describes the basics of 2D cellular automata and the 2D CA concepts used for our encryption scheme. Section 3 presents our new non-parametric test for pixel randomness. Section 4 describes our 2D CA based encryption scheme. Section 5 presents the simulation results and performance evaluations. Section 6 gives the concluding remarks.

\section{2D Cellular Automata}

The extensional behavior of 1D Cellular Automata is also able to produce complicated patterns in two-dimension. This extension is significant since this is compared with pattern formation in physical systems. 2D CA is a regular 2D lattice of cells. Each cell has n possible values and is updated in each discrete time steps according to a rule f that depends on the value of sites in some neighbor around it. There are different types of lattices and neighborhood structures in a 2D CA. Figure 1 shows the two familiar neighborhood structures named as von-Neumann neighborhood and Moore neighborhood. The value of $a_{ij}$ of a cell at $(i,j)$th position in $(t+1)$th time in a 2D CA with a rule f, that depends only on the cells according to von-Neumann neighborhood is evolved from $a_{i,j}^{(t+1)} = f(a_{i,j}^{t},a_{i,j+1}^{t},a_{i+1,j}^{t},a_{i,j-1}^{t},a_{i-1,j}^{t}) $.\\
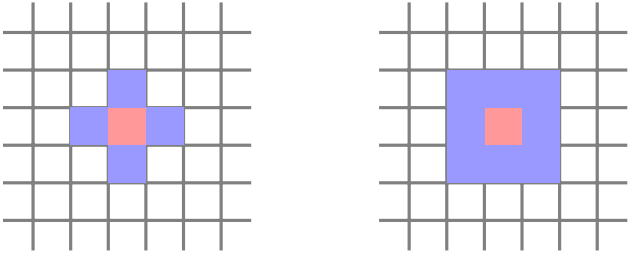
\begin{figure}
\begin{tikzpicture}
\large ({\bf a})
\draw[step=0.5cm,gray,very thick] (-1.9,-1.9) grid (1.4,1.4);
\fill[red!40!white] (-0.5,-0.5) rectangle (0,0);
\fill[blue!40!white] (-0.5,0) rectangle (0,0.5);
\fill[blue!40!white] (0,-0.5) rectangle (0.5,0);
\fill[blue!40!white] (-0.5,-1) rectangle (0,-0.5);
\fill[blue!40!white] (-1,-0.5) rectangle (-0.5,0);
\hspace{1cm}
\large ({\bf b})
\draw[step=0.5cm,gray,very thick] (2.1,-1.9) grid (5.4,1.4);
\fill[red!40!white] (3.5,-0.5) rectangle (4,0);
\fill[blue!40!white] (3,0) rectangle (4.5,0.5);
\fill[blue!40!white] (3,-1) rectangle (3.5,0);
\fill[blue!40!white] (3.5,-1) rectangle (4.5,-0.5);
\fill[blue!40!white] (4,-0.5) rectangle (4.5,0);
\end{tikzpicture}\\
 \caption{{von-Neumann and Moore Neighborhoods}   }         
\end{figure}                  
\subsection{Mathematical Model}

1D and 2D CA are studied differently with the help of polynomial algebra and matrix algebra by researchers. P.P.Choudary et.al.already have studied and designed a new characteristic of 2D CA\cite{chou2,chou1}. These concepts are used for our proposed encryption method.
	2D nine neighborhood Moore’s neighborhood is used here for evolutions. Different rules are defined in terms of dependencies from the following rule convention.\\
	
	\renewcommand{\arraystretch}{1.5}
	{
	\hspace{6cm}
	\begin{tabular}{|c|c|c|l|}
\hline
64 &  128 & 256 \\  \hline  
32 &    1   &   2   \\ \hline   
16 &    8   &   4   \\ \hline

\end{tabular}
}
       
Here the central box is the cell being considered for evolution and the other boxes are its neighbors. The numbers in the boxes are used to specify the boxes for a particular rule. For example, rule 2 characterizes the dependency on its right neighbor whereas such dependency only on its bottom neighbor is characterized by rule 8 and so on. Rules, which are having dependency only on one cell are called fundamental rules. Suppose a cell has dependency on more than 2 neighbors. Then the rule number will be the sum of the numbers of the corresponding cells. For example 2D CA rule 15 (1+2+4+8) refers to the bottom-right 2x2 cells. There are 512 such rules including zero dependency.

These 512 rules are classified in to 9 different groups in terms of the number of 1’s in the binary representation of the rule number. For example rule 8 belongs to Group 1, rule 320 belongs to Group 2 etc.,
	
\subsection{Uniform CA Rules}
Evolution of such 2D CA rules can be considered as matrix operations with 0’s and 1’s. If same rule is applied to each entry of the given problem matrix, then the CA is called uniform CA. Example 2.1 shows the application of the rule 15 in the problem matrix of size (2x3) with null boundary condition.

{\bf Example 2.1:}
\[ \left( \begin{array}{ccc}
1 & 0 & 0 \\
0 & 1 & 0 \end{array} \right)
\xrightarrow{Rule 15}
 \left( \begin{array}{ccc}
0 & 1 & 0 \\
1 & 1 & 0 \end{array} \right)\]
Let $S_{t}$ denote the binary matrix which represents the current state of a 2D CA. The next state of a cell is obtained by calculating XOR operation of the states in the corresponding boxes of a rule. In this example the states in the dependent neighbors of the 1st cell are 1,0,0,1 according to rule 15 and its XOR value 0 is its next state value.\\

	To apply the rules to the problem matrix, one can use the rule matrices which are used in \cite{pries}. For a problem matrix of size (m x n), the size of the rule matrix $R_{n}$ is (mn x mn) since the problem matrix is applied as a vector of order (mn x 1) got from the row major order.\\\\
{\bf Structure of rule matrices:}\\\\
	Let $R_{n}$ denote the rule matrix for rule n, for n=$0,1,2,...,511$. It is found that these rule matrices are related to each other in the following way.\\
$R_{1}$ $\rightarrow$ Identity matrix.\\
$(R_{2})^{T} = R_{32}, (R_{4})^{T} = R_{64}, (R_{8})^{T} = R_{128}, (R_{16})^{T} = R_{256}$ ,\\
 where the power T stands for the transpose of the matrix. By having these 5 fundamental matrices, one can easily get the rule matrices of all other rules using transpose and the XOR operations as given in example 2.2.

{\bf Example 2.2:}\\\\
$R_{15} = R_{8} + R_{4} + R_{2} + R_{1}$.\\
$R_{82} = R_{64} + R_{16} + R_{2} = (R_{4})^{T} + R_{16} + R_{2}$.\\
$R_{327} = R_{256} + R_{64} + R_{4} + R_{2} + R_{1} = (R_{16})^{T} + R_{64} + R_{4} + R_{2} + R_{1}$.\\

	In \cite{chou4} P.P.Choudary et. al. have found new way of constructing rule matrices using the following two binary sequences $S_{1}$ and $S_{2}$.\\
$S_{1}$ = $\underbrace{111..1}0\underbrace{111..1}0\underbrace{111..1}0..0\underbrace{111..1}$\\
That is $S_{1}$ starts with (n-1) $1^{'}s$ followed by one 0 then (n-1) $1^{'}s$ followed by one 0 and finally ends with (n-1) $1^{'}s$.\\
 $S_{2}$ = $0\underbrace{111..1}0\underbrace{111..1}0\underbrace{111..1}0..0\underbrace{111..1}0$\\
That is $S_{2}$ starts with one 0 followed by (n-1) $1^{'}s$ and one 0 then (n-1) $1^{'}s$ and ends with (n-1) $1^{'}s$ followed by one 0, where n is the number of columns in the problem matrix.\\

The five basic matrices are defined as follows with the structure of above sequences.\\
$R_{1}$ = Identity matrix.\\
$R_{2}$ = (n-1)th diagonal elements are arranged like sequence $S_{1}$ and all other elements are 0.\\
$R_{4}$ = (n+1)th diagonal elements are arranged like sequence $S_{1}$ and all other elements are 0.\\
$R_{8}$ = nth diagonal elements are 1 and all other elements are 0.\\
$R_{16}$ = (n-1)th diagonal elements are arranged like sequence $S_{2}$ and all other elements are 0.\\
Other rule matrices are defined as in example 2.
	Among all these 512 rules we are interested only in rules with non-singular  matrices as they are reversible which are useful in Cryptography. There are only 31 rules that are reversible and those rules are expressed in terms of $R_{1}$ since they have $1^{'}s$ in their diagonal.
\subsection{Hybrid CA Rules}

Unlike uniform CA, in hybrid CA each cell has its own local rule, that is, different rule is used for different cell. The total number of matrices of order (mn x mn) with entries 0 and 1 is ${2^{(mn)}}^{2}$, whereas we dealt with only 512 rules which are corresponding to linear Boolean rules. The rule matrices of hybrid rules we are discussing here are corresponding to remaining ${2^{(mn)}}^{2}$-512 matrices.\\\\\\\\
{\bf Example 2.3:}\\\\
\begin{picture}(20,10)
\put(225,-159){\vector(2,-1){30}}
\put(225,-136){\vector(1,0){30}}
\put(205,-133){\vector(3,1){53}}
\end{picture}
	Let us consider the hybrid rule which defines as follows for a 2x2 problem matrix.\\
1.	1st cell changes its state by looking at the values in 1st, 3rd and 4th positions of the problem matrix.\\
2.	2nd cell changes its state by looking at the values in 2nd and 3rd positions of the problem matrix.\\
3.	3rd cell changes its state by looking at the values in the 4th position of the problem matrix.\\
4.	4th cell changes its state by looking at the values in the 1st, 2nd and 3rd position of the problem matrix.\\
\[ \left( \begin{array}{cc}
a & b \\
c & d \end{array} \right)
\left( \begin{array}{cccc}
1 & 0 & 1 & 1 \\
0 & 1 & 1 & 0 \\
0 & 0 & 0 & 1 \\
1 & 1 & 1 & 0 \end{array} \right)\]
All such non-singular matrices of order (mn x mn) are reversible and they can be used for cryptography purposes.

\section{A new Pixel Randomness Algorithm}
Strength of an image encryption method measured by showing the randomness in the pixels of the shuffled and encrypted images. In literature we already have some parametric tests, which are used for checking randomness in the pixels of the images. But these tests are based on the assumption that the observations are treated as samples drawn from normal populations. If this is the case, these methods extract all the information that is available in a sample, and they usually attain the best possible precision. However, since there are many situations where it is doubtful whether the assumption of normality can be met, so the non-parametric tests came in to picture. It is useful to have a technique for testing whether a sample may be looked upon as random after it has been obtained. One such technique is based on the order in which the sample values were obtained, more specifically, it is based on the number of runs exhibited in the sample results\cite{gib1}.\\

A process which produces independent and identically distributed (i.i.d) samples is called a "truly random process" in Statistics. A process is not considered truly random if the value in the sequence can be determined by its position. The property of iid is equivalent to the randomness, and it forms the basis for many statistical tests. Many practical applications will not have true random observations, but these simple tests are still useful to check how far the observations being close to randomness. It has very good application in cryptographic security, where it is used to check the randomness in various random number generators.\\

Investigations of randomness of a given sequence often require statistical tools for distribution comparison. Among them, goodness-of-fit tests and entropy estimates are two well-understood concepts\cite{weg}. However, when the distribution of the observed data is unknown, , we have to go for non parametric tests, using some distribution invariant properties of random process. For example, the observations are transformed to a sequence of symbols that can specify their relative positions or magnitudes. The pattern of this sequence will serve as a measure of the randomness of the original process.\\

A run is defined as a succession of one or more identical symbols in the ordered symbol sequence. For example, the 0’s and 1’s in a bit string such as 0000011111 and 0011001101 have 2 and 6 runs respectively. The number of runs and their sizes are used as a measure of the randomness of the observations. In a truly random sequence, we very rarely encounter too few, too many runs or runs with long length, etc., So the presence of any of these in an ordered symbol sequence is used as a statistical criteria for rejecting null hypothesis. Also these criteria are related to each other as too few runs results in some long length  runs and too many runs results in some short length runs. So the number of runs is considered as a useful parameter in the randomness test.\\

Dichotomizing criteria is used for symbolizing the sequence for quantitative observations. By comparing each number in the sequence to a focal point (mean or median), it can be symbolized as + or – depending on whether the number is greater or lesser than the focal point. Also, the trend or autocorrelation of the sequence can be determined from the relative magnitudes or ranks of adjacent numbers.\\

If a sequence contains $n_{1}$ symbols of one kind and $n_{2}$ of another kind, the sampling distribution of the total number of runs, $u$, can be approximated closely by a normal distribution with\\
	\begin{equation}
	\mu_{u}=\frac{2n_{1}n_{2}}{n_{1}+n_{2}}+1
	\hspace{1cm} and
	\hspace{1cm}
	 \sigma_{u}=\sqrt{\frac{2n_{1}n_{2}(2n_{1}n_{2}-n_{1}-n_{2})}{(n_{1}+n_{2})^{2}(n_{1}+n_{2}-1)}} 
	\end{equation}
	
	Thus the test of the null hypothesis $H_0$that the arrangement of the symbols is random, can be based on the statistic 
	\begin{equation}
	Z=\frac{u-\mu_{u}}{\sigma_{u}}
	\end{equation}
	which has approximately the standard normal distribution.\\
	In order to carry out the test for randomness, we set $H_0$ as arrangement is random against the alternating hypothesis $H_1$ as arrangement is not random. By setting the level of significance as $\alpha$, we make use of the following criterion for the critical region. If $z<-z_{\alpha/2}$ or $z>z_{\alpha/2}$ then reject $H_0$ and then we conclude the result as arrangement is random or not with $(1-\alpha) \times 100\% $level of confidence. We take $\alpha = 0.01$ and make the conclusion about randomness with 99\% of confidence. One can see from the standard normal distribution table that $z_{\alpha/2}=2.575$.\\
	
	In our method the runs are formed in the following manner: First, t distinct pixels are selected randomly from the image for which the pixel randomness is to checked. Now for each of these t pixels consider its Moore's neighborhood pixels including the center pixel. Find the mean value of these 9 pixels for all the t pixels. Now we have t number of mean values. Now move horizontally(or vertically or diagonally) two steps from each of these t pixels, again we have now t points and find the t number of Moore's neighborhood mean values for these t pixels. Now we have totally 2t mean values. Do the same process h number of times, and we will have 'ht' mean values. Find the median for all these ht mean values. Compare this median value with each of these ht mean values, if the mean value is less than the median then make the run entry as $a$ otherwise $b$. Now we will have a sequence of $a^{'}s$ and $b^{'}s$. This sequence is now exhibiting the runs. With these runs, the test has to be done using the equations 1 and 2. Moving along horizontal(or vertical or diagonal) direction is required to consider the distribution of pixels focally. The following algorithm explains the test.\\
	\begin{small}
	\begin{spacing}{1}
	{\bf Algorithm:}\\
	Notations:\vspace{-0.2cm}
\begin{itemize}
\itemsep-0.5em
\item P is an image to be tested.
\item N is the number of tests.
\item T is the number of evaluations.
\item n is the number of random pixels from P.
\item h is the number of horizontal or vertical movements.
\end{itemize}
1.\hspace{1cm}$l=1$.\\
2.\hspace{1cm}for $i$=1 to T\\
3.\hspace{1.5cm}$c_1=0$\\
4.\hspace{1.5cm}for $j$=1 to N\\
5.\hspace{2cm}Choose n random pixels $(p_1,p_2,p_3,..p_n)$without repetition.\\
6.\hspace{2cm}for $t$=1 to $h$\\
7.\hspace{2.5cm}Compute n number of means $(m_{t1},m_{t2},m_{t3},..m_{tn})$from each pixel's Moore's neighborhood.\\
8.\hspace{2.5cm}Move 2 pixels horizontally(or vertically or diagonally).\\
9.\hspace{2cm}end for\\
10.\hspace{2cm}Find the median $d$ for the $hn$ means.\\
11.\hspace{2cm}for $s$=1 to $h$\\
12.\hspace{2.5cm}for $t$=1 to $n$\\
13.\hspace{3cm}if($m_{st}>d$)\\
14.\hspace{3.5cm}$r_{st}$=1.\\
15.\hspace{3cm}else\\
16.\hspace{3.5cm}$r_{st}$=0.\\
17.\hspace{3cm}end if\\
18.\hspace{2.5cm}end for\\
19.\hspace{2cm}end for\\
20.\hspace{2cm}Count the number of 1's and 0's in $r$ and denote as $n_1$ and $n_2$ respectively.\\
21.\hspace{2cm}Count the number of runs $u$ in $r$.\\
22.\hspace{2cm}Compute $\mu_{u}$ and $\sigma_{u}$ using 1.\\
23.\hspace{2cm}Perform the test using the statistic 2.\\
24.\hspace{2cm}if ($-z_{\alpha/2}<z<z_{\alpha/2}$) \\
25.\hspace{2.5cm}$c_1=c_1+1$.\\
26.\hspace{2cm}end if\\
27.\hspace{1.5cm}end for\\
28.\hspace{1.5cm}$c_2[l]$=$c_1$.\\
29.\hspace{1.5cm}$l=l+1$.\\
30.\hspace{1cm}end for\\
31.\hspace{1cm}C=(($\sum_{k=1}^{T}c_2[k])/T)*100$.\\
C measures the image pixel randomness in the range 1-100.\\
\end{spacing}
\end{small}
   
\section{Encryption Scheme}

The new 2D CA based encryption scheme in this paper is considered for gray scale and color images. Basic idea of the proposed image encryption, decryption method is based on the idea in \cite{chou4}. Pixel values are changed by considering its binary representation using 2D CA. Encryption process has been done in two parts. In the first part, image scrambling is performed on the binary string form of the image and in the second part this scrambled image is been encrypted using 2D CA by considering the problem matrix of size $(5$ x $5)$.
\vspace{-0.5cm}
\subsection{Encryption Algorithm}
This scheme can be applied on any square size problem matrix. We have implemented the scheme for the problem matrix of size $(5$ x $5)$.If the plain image size is mxn then the size of the plain image in binary form is mx8n. Pad rows and columns with 0’s in the binary form such that the number of rows and columns are multiple of 5. Let the size of this matrix be $(p_{1}$ x $p_{2})$. Perform the permutation on the columns and rows of this matrix. This scrambled image is now considered as the input image for the 2D CA encryption scheme.
Extract the $(5$ x $5)$ sub matrices in row major order from the scrambled binary image of size $(p_{1}$ x $p_{2})$. Now we have $(\frac{p_{1}}{5}$ x $\frac{p_{2}}{5})$ array P of $(5$ x $5)$ sub matrices. Consider each $(5$ x $5)$ sub matrix as a problem matrix and apply the proposed 2D CA scheme. Rules with invertible rule matrices are used for encryption purpose.\\
Encryption can be done in the following different ways:\\
W1:	All $(5$ x $5)$ problem matrices in P use the same invertible uniform rule.\\
W2:	Each row (column) of P uses different invertible uniform rules.\\
W3:	Each problem matrix in P uses a different invertible uniform rule.\\
W4:	All $(5$ x $5)$ problem matrices in P use the same invertible hybrid rule.\\
W5:	Each row (column) of P uses different invertible hybrid rules.\\
W6:	Each problem matrix in P uses a different invertible hybrid rule.\\
As we move from W1 to W6, the size of the key space is increased. After following any one from the above 6 ways, we will get the encrypted image. For the RGB color image, the encryption has to be done on each component separately. For the sake of increasing the difficulty of cryptanalysis, one can always follow W3 or W6 for encryption. The following algorithm explains the encryption process in W1.
\begin{small}
	\begin{spacing}{1}
	{\bf Algorithm:}\\
	Get the matrix P from the binary scrambled plain image by clubbing (5x5) submatrices in row major order.\\
	{\bf Input:} Matrix P of size (mxn) of (5x5) submatrices.\\
	{\bf Output:}Matrix C of size (mxn) of size(5x5) submatrices.\\
	1.\hspace{1cm}Choose an invertible uniform rule r.\\
	2.\hspace{1cm}Construct the rule matrix R of size (25x25).\\
	3.\hspace{1cm}for $i$=1 to m\\
	4.\hspace{1.5cm}for $j$=1 to n\\
	5.\hspace{2cm}Get the vector $P^\prime(i,j)$ of size (25x1) from $P(i,j)$ by ordering the elements in row major order.\\
	6.\hspace{2cm}$C^\prime(i,j)=R*P^\prime(i,j).$\hspace{1cm}(where * is matrix multiplication in $Z_2$).\\
	7.\hspace{2cm}Get the matrix $C(i,j)$ of size (5x5) from $C^\prime(i,j)$  by arranging the elements in row major order.\\
	8.\hspace{1.5cm}end for\\
	9.\hspace{1cm}end for\\
	\end{spacing}
\end{small}
\vspace{-0.3cm}
We will get the encrypted image from the gray scale representation of the re-ordered binary matrix C.
\subsection{Decryption}

As we have used invertible uniform and hybrid rules for encryption, inverse exists for all the encryption rules. So the decryption can be done using the inverse rule matrices of the encryption rules matrices. For the color images, decryption can be done on the three components separately using the inverse rule matrix (matrices) of the corresponding encryption rule matrix (matrices).
\section{Simulation Results and Performance evaluation}
Several simulations are conducted to test various properties of the proposed 2D CA image encryption/decryption scheme including information concealing, confusion and diffusion properties.
For the gray scale pepper image and the RGB color Lena images, we have performed all the ways of encryption which are mentioned above. Figure 2 shows the experimental results of the encryption scheme. (a) shows the gray scale Pepper image, (b),(c) and (d) show the encrypted images of (a) by W1, W2 and W3 respectively. (e) shows the RGB color Lena image, (f), (g) and (h) show the encrypted images of (e) by W4, W5 and W6 respectively. 
\begin{figure}[h!]
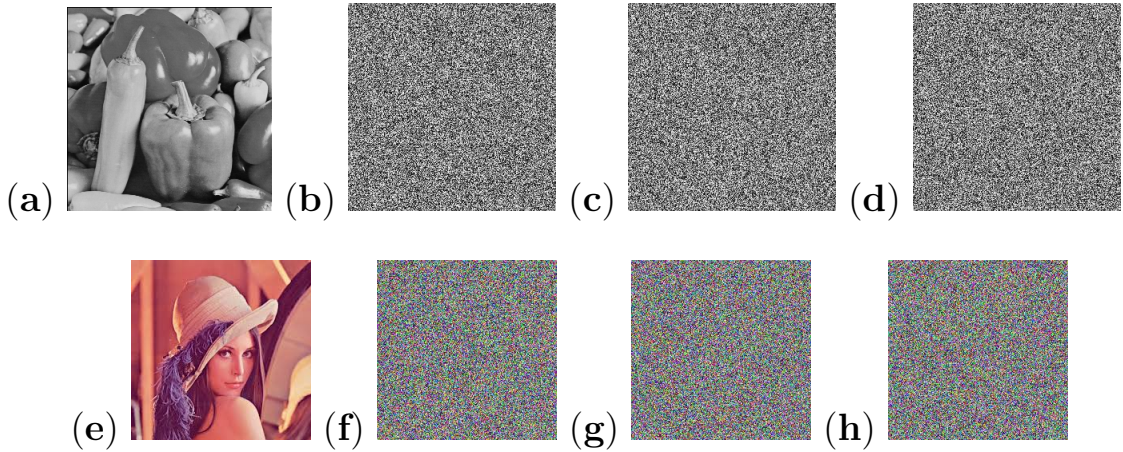

\large ({\bf a})
\centering
\includegraphics[scale = 0.3]{peppers256.pdf}
\large ({\bf b})
\includegraphics[scale = 0.3]{peppers256_enc_w1.pdf}
\large ({\bf c})
\includegraphics[scale = 0.3]{peppers256_enc_w2.pdf}
\large ({\bf d})
\includegraphics[scale = 0.3]{peppers256_enc_w3.pdf}
\vspace{0.5cm}

\large ({\bf e})
\includegraphics[scale = 0.3]{lena.pdf}
\large ({\bf f})
\includegraphics[scale = 0.3]{Encrypted_Lena_w4.pdf}
\large ({\bf g})
\includegraphics[scale = 0.3]{Encrypted_Lena_w5.pdf}
\large ({\bf h})
\includegraphics[scale = 0.3]{encrypted_Lena_w6.pdf}
\begin{center}
\caption{{\bf(a)} Original Pepper Image; {\bf(b)} Encrypted (a) by W1; {\bf(c)} Encrypted (a) by W2; {\bf(d)} Encrypted (a) by W3; {\bf(e)} Original Lena Image; {\bf(f)} Encrypted (e) by W4; {\bf(g)} Encrypted (e) by W5; {\bf(h)} Encrypted (e) by W6.}
\label{overflow}
\end{center}
\end{figure}

\subsection{Information concealing}

If the key in the receiver side is identical with the secret key, the image can be revealed without loss. Figure 3 shows the results of our experiments with gray scale House image and the Lena color image, in both the experiments we could get back the original images using the correct secret key. In the figure 3, (a) and (d) are the original images, (b) and (e) are the encrypted images and (c) and (f) are the decrypted images.
\begin{figure}[h!]
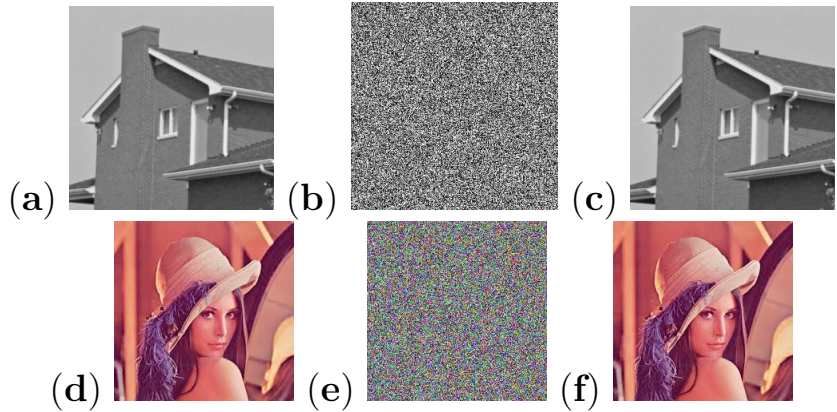

\large ({\bf a})
\centering
\includegraphics[scale = 0.3]{house.pdf}
\large ({\bf b})
\includegraphics[scale = 0.3]{house_enc.pdf}
\large ({\bf c})
\includegraphics[scale = 0.3]{house_dec.pdf}
\hspace{8cm}
\large ({\bf d})
\includegraphics[scale = 0.3]{lena3.pdf}
\large ({\bf e})
\includegraphics[scale = 0.3]{lena3_enc.pdf}
\large ({\bf f})
\includegraphics[scale = 0.3]{lena3_dec.pdf}
\caption{{\bf(a)} Original House Image; {\bf(b)} Encrypted House Image by W3; {\bf(c)} Decrypted House Image; {\bf(d)} Original Lena Image; {\bf(e)} Encrypted Lena Image by W1; {\bf(f)} Decrypted Lena Image.}
\label{overflow}
\end{figure}

\subsection{Confusion Property}

It is always essential to make the relationship between the key and cipher text as complex as possible so that the attackers can not get the key from the cipher text. Histograms of the original and encrypted images of the above experiments are given in Figure 4. The histograms of the encrypted images are almost uniform regardless of the original images and are significantly different from those of the original images. The results exhibit the confusion property of the proposed scheme.
\begin{figure}[h!]
\large ({\bf a})
\centering
\includegraphics[scale = 0.3]{peppers256.pdf}
\includegraphics[scale = 0.2]{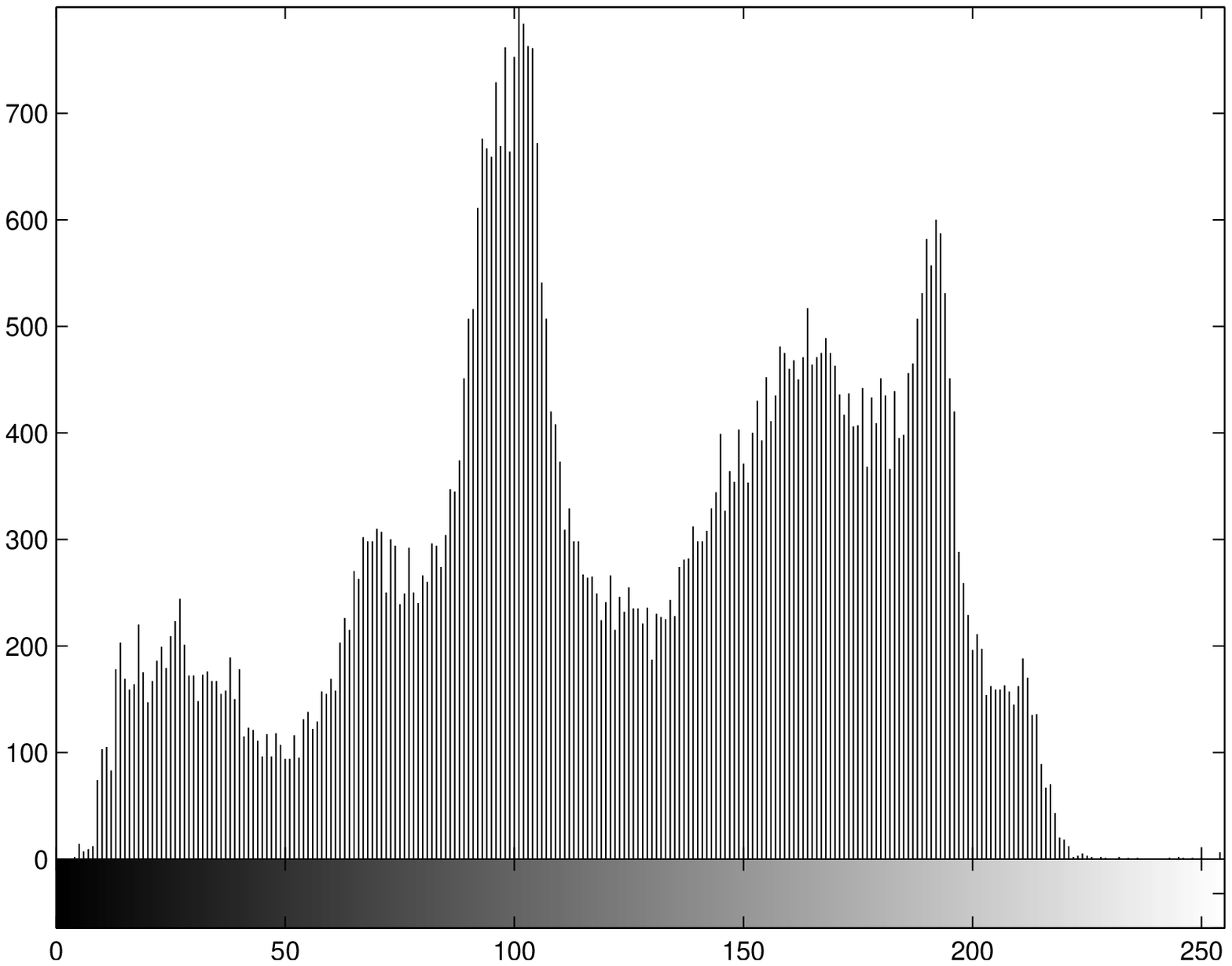}
\hspace{1cm}
\large ({\bf b})
\includegraphics[scale = 0.3]{peppers256_encrypt1_with_481.pdf}
\includegraphics[scale = 0.2]{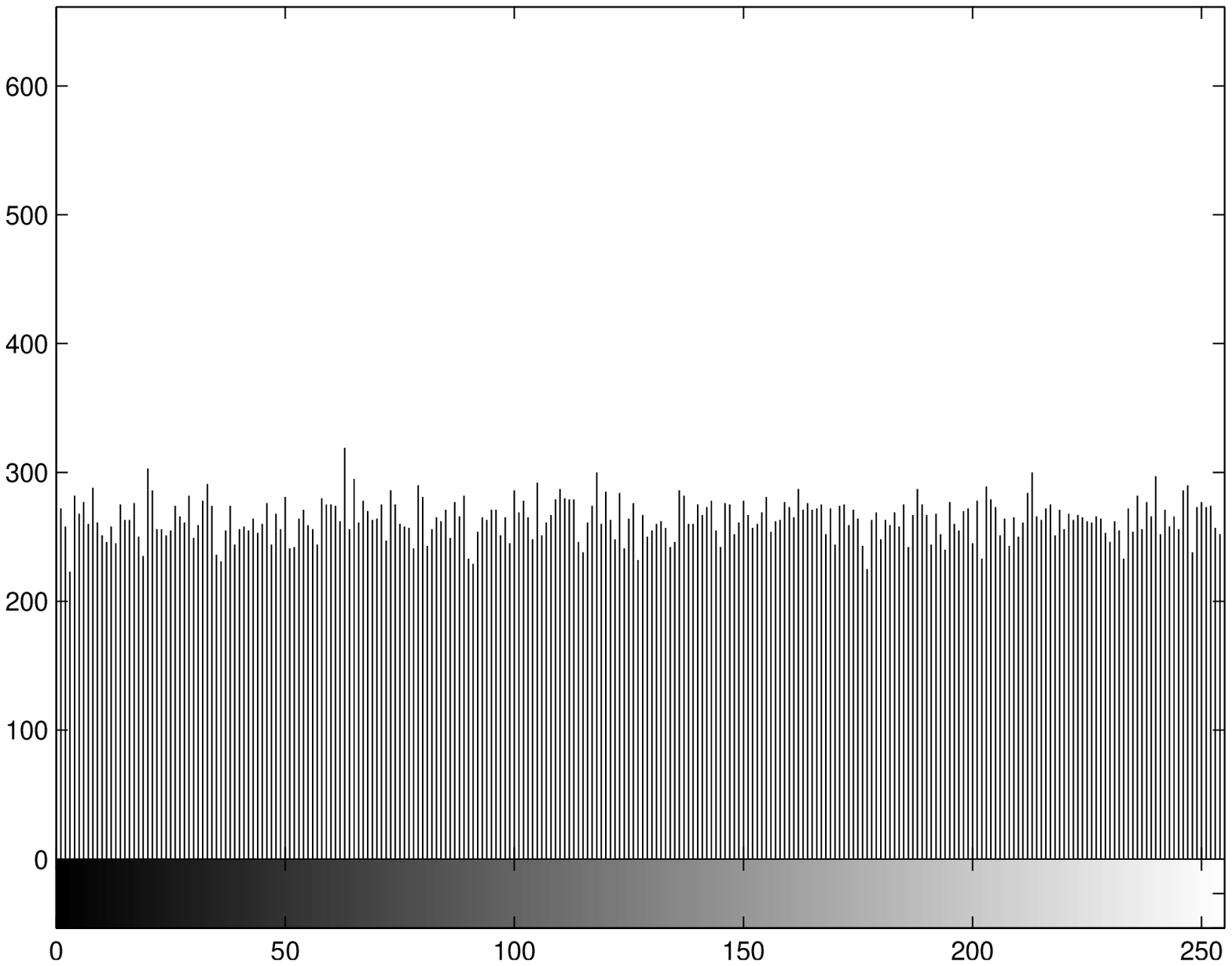}

\large ({\bf c})
\includegraphics[scale = 0.15]{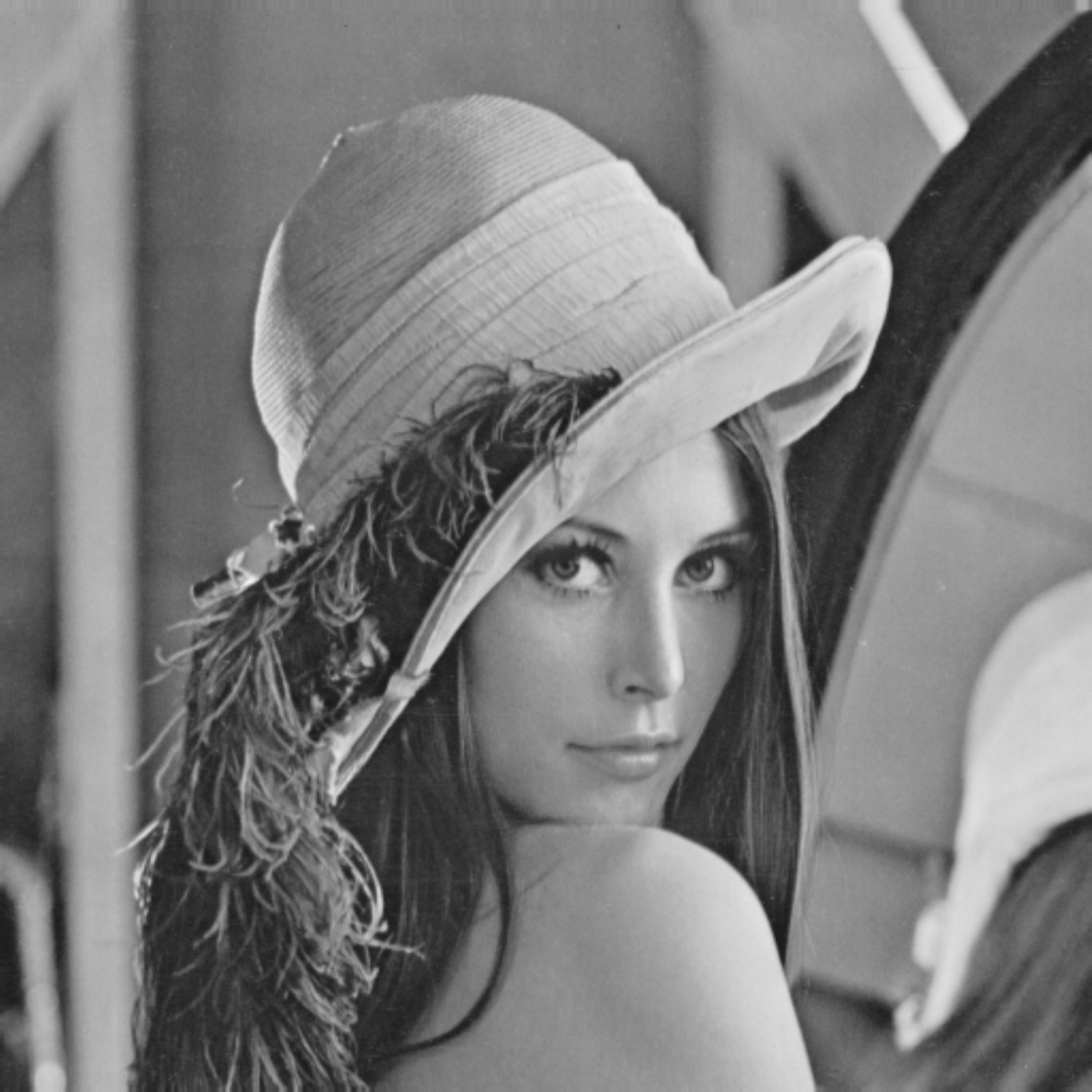}
\includegraphics[scale = 0.2]{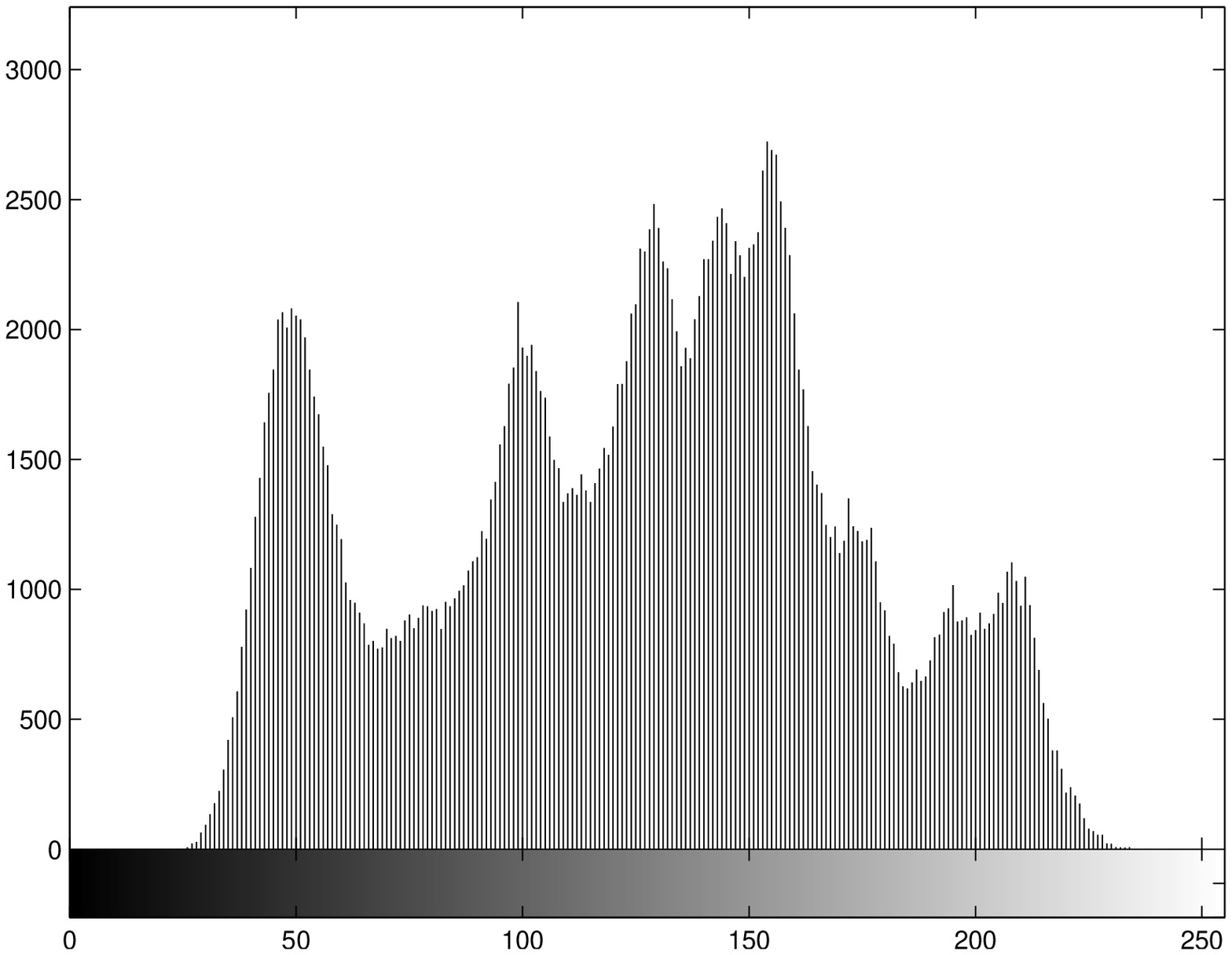}
\hspace{1cm}
\large ({\bf d})
\includegraphics[scale = 0.15]{lena_encrypt1_with_225.pdf}
\includegraphics[scale = 0.2]{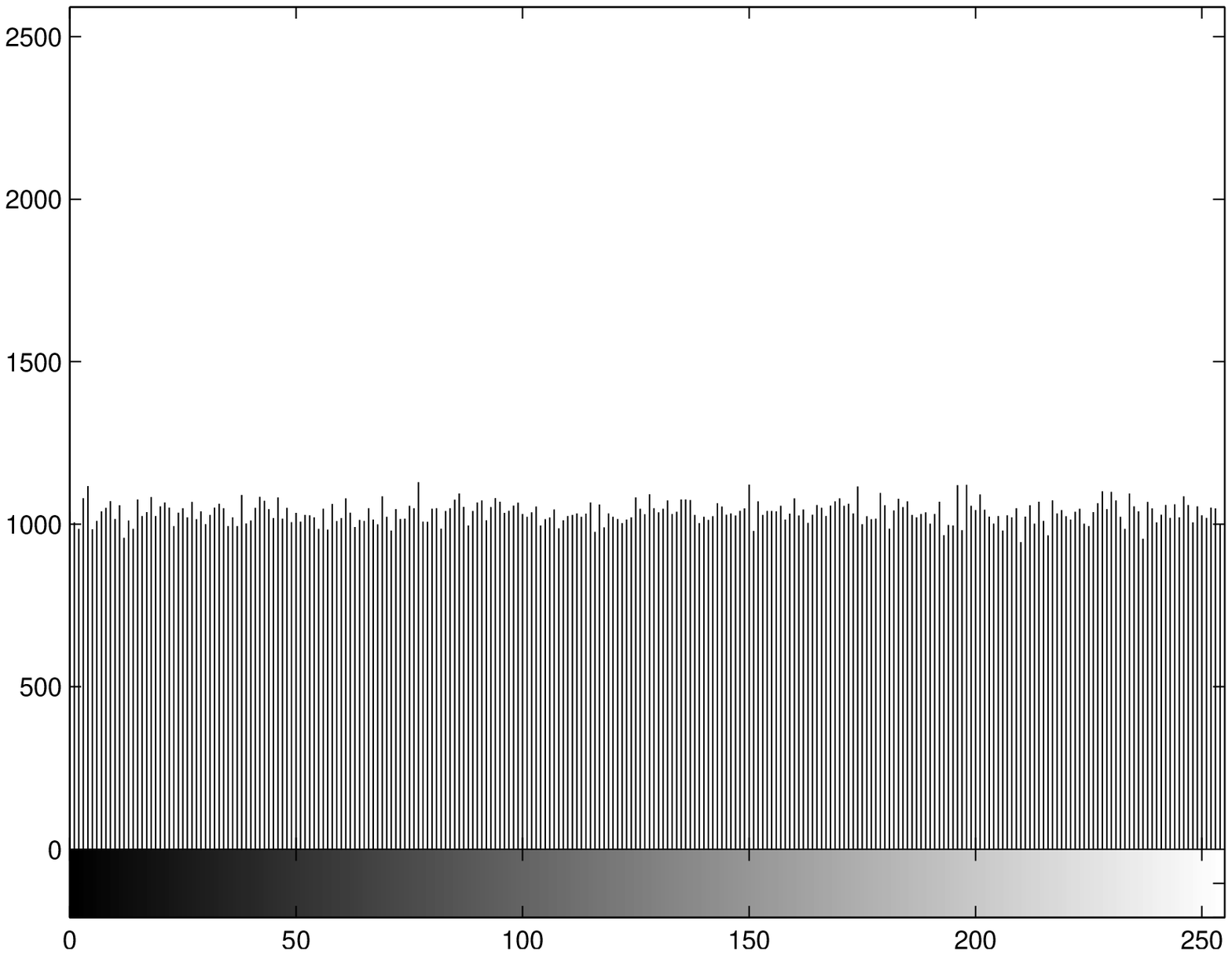}

\caption{{\bf(a)} Original Pepper Image and its Histogram; {\bf(b)} Encrypted Pepper Image and its Histogram.\\{\bf(c)} Original Lena Image and its Histogram; {\bf(d)} Encrypted Lena Image and its Histogram.}
\label{overflow}
\end{figure}

\subsection{Diffusion Property}
In any strong cryptosystem, key plays the main role to spread out the redundant bits in the plain image over the cipher image along with the encryption method. So small changes in the key, produces a significantly different cipher image. This property is termed as diffusion property. We have performed encryption and decryption key sensitivity tests to show the diffusion property in our method. Figures 5 \& 6 show the test results of encryption and decryption key sensitivity analysis respectively. These results clearly exhibit the diffusion property.
\begin{center}
\begin{figure}[h!]
\large ({\bf a})
\centering
\includegraphics[scale = 0.3]{peppers256.pdf}
\large ({\bf b})
\includegraphics[scale = 0.3]{ekeyanaly1.pdf}
\large ({\bf c})
\includegraphics[scale = 0.3]{ekeyanaly2.pdf}
\large ({\bf d})
\includegraphics[scale = 0.3]{ekeyanaly3.pdf}
\caption{Encryption Key sensitivity\\{\bf(a)} Original pepper Image; {\bf(b)} Encrypted with K1; {\bf(c)} Encrypted with K2; {\bf(d)} Difference of b \& c; }
\label{overflow}
\end{figure}
\end{center}
\begin{center}
\begin{figure}[h!]
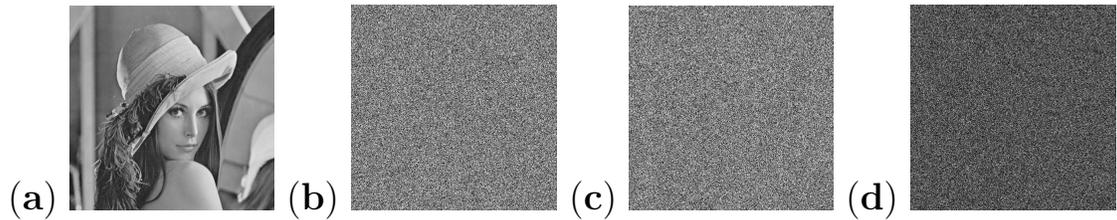

\large ({\bf a})
\centering
\includegraphics[scale = 0.15]{lena_gray.pdf}
\large ({\bf b})
\includegraphics[scale = 0.15]{lena_encrypt1_with_225.pdf}
\large ({\bf c})
\includegraphics[scale = 0.15]{lena_enc_with_225_dec_with_481.pdf}
\large ({\bf d})
\includegraphics[scale = 0.15]{lena_enc_dec_diff_with_481_and_225.pdf}

\caption{ Decryption Key sensitivity \\ {\bf(a)} Original Lena Image; {\bf(b)} Encrypted with K1; {\bf(c)} Decrypted with K2; {\bf(d)} Difference of b \& c.}

\label{overflow}
\end{figure}
\end{center}
\subsection{Statistical Analysis}
\subsubsection{Correlation Analysis}
In a perfectly encrypted images and in a perfectly shuffled images, any pair of adjacent pixels along horizontal, verical or diagonal directions should not be correlated. 1000 random pairs of adjacent pixels have been chosen horizontally, vertically and diagonally from the plain and encrypted images for the correlation analysis. The correlation coefficient has been computed between two adjacent pixels using the equation 6.


\begin{equation}
E(x) = \frac{1}{N} \sum_{i=1}^{N} x_{i}
\end{equation}

\begin{equation}
D(x) = \frac{1}{N} \sum_{i=1}^{N} ((x_{i} - E(x))^{2}
\end{equation}

\begin{equation}
cov (x, y) = \frac{1}{N} \sum_{i=1}^{N}  (x_{i} - E(x))(y_{i} - E(y))
\end{equation}

\begin{equation}
\gamma_{xy} = \frac{ cov (x, y)  }{\sqrt{D(x)}  \sqrt{D(y)}	}
\end{equation}
Table [1] gives the correlation coefficient values of adjacent pixels in the Lena and Pepper images along horizontal, vertical and diagonal directions of the plain and encrypted images. This confirms that the adjacent pixels in the plain images are strongly correlated where as the adjacent pixels in the encrypted images are weakly correlated. Figure 7 illustrates the correlation distribution of the plain and encrypted Lena, Pepper as well as House images. Correlations along horizontal, vertical and diagonal directions are tested for Lena, Pepper and House images respectively. We can clearly see that the encrypted images are very weakly correlated.

\begin{center}
\begin{tabular}{|c|c|c|c|c|l|}
\hline
\centering
 Images &  Horizontal  & Vertical  & Diagonal   \\  \hline  
Original Lena Image  & 0.9746 & 0.9381 & 0.9438 \\ \hline   
Encrypted Lena Image &  -0.0184  & 0.0352 & 0.0228 \\ \hline
Original Pepper  Image &  0.9143  & 0.9403 & 0.9552 \\  \hline
Encrypted Pepper  Image &  0.0110  & 0.0065 & -0.0104 \\ \hline
\end{tabular}\\
{\bf Table 1:}  Correlation coefficients between adjacent pixels for original and encrypted images.
\end{center}

\begin{figure}[h!]
\hspace{4cm} {\bf Original} \hspace{0.8cm}      {\bf  Encrypted} \\
\large ({\bf a})
\centering
\includegraphics[scale = 0.15]{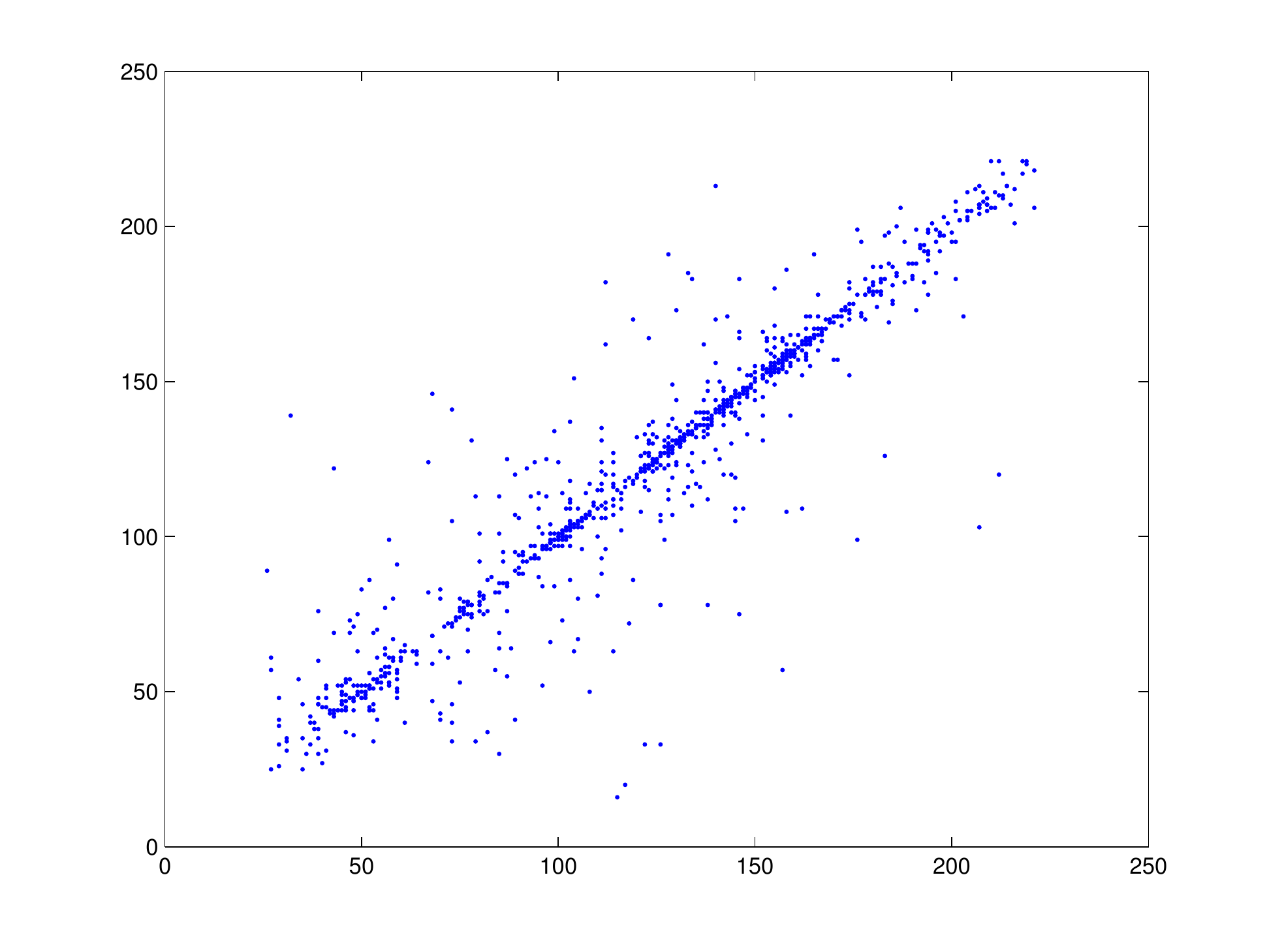}
\includegraphics[scale = 0.15]{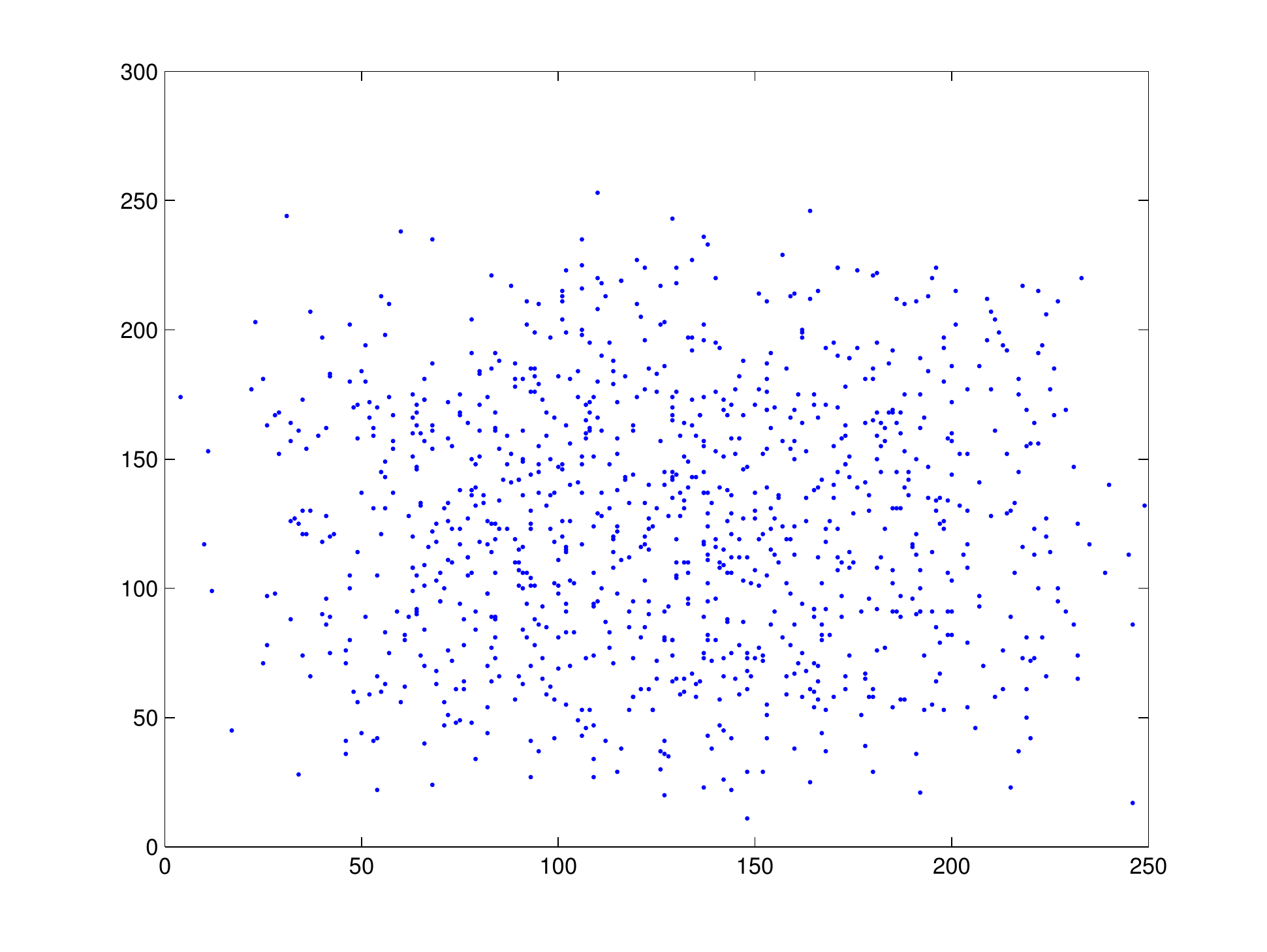} \\
\vspace{1cm}
\large ({\bf b})
\includegraphics[scale = 0.15]{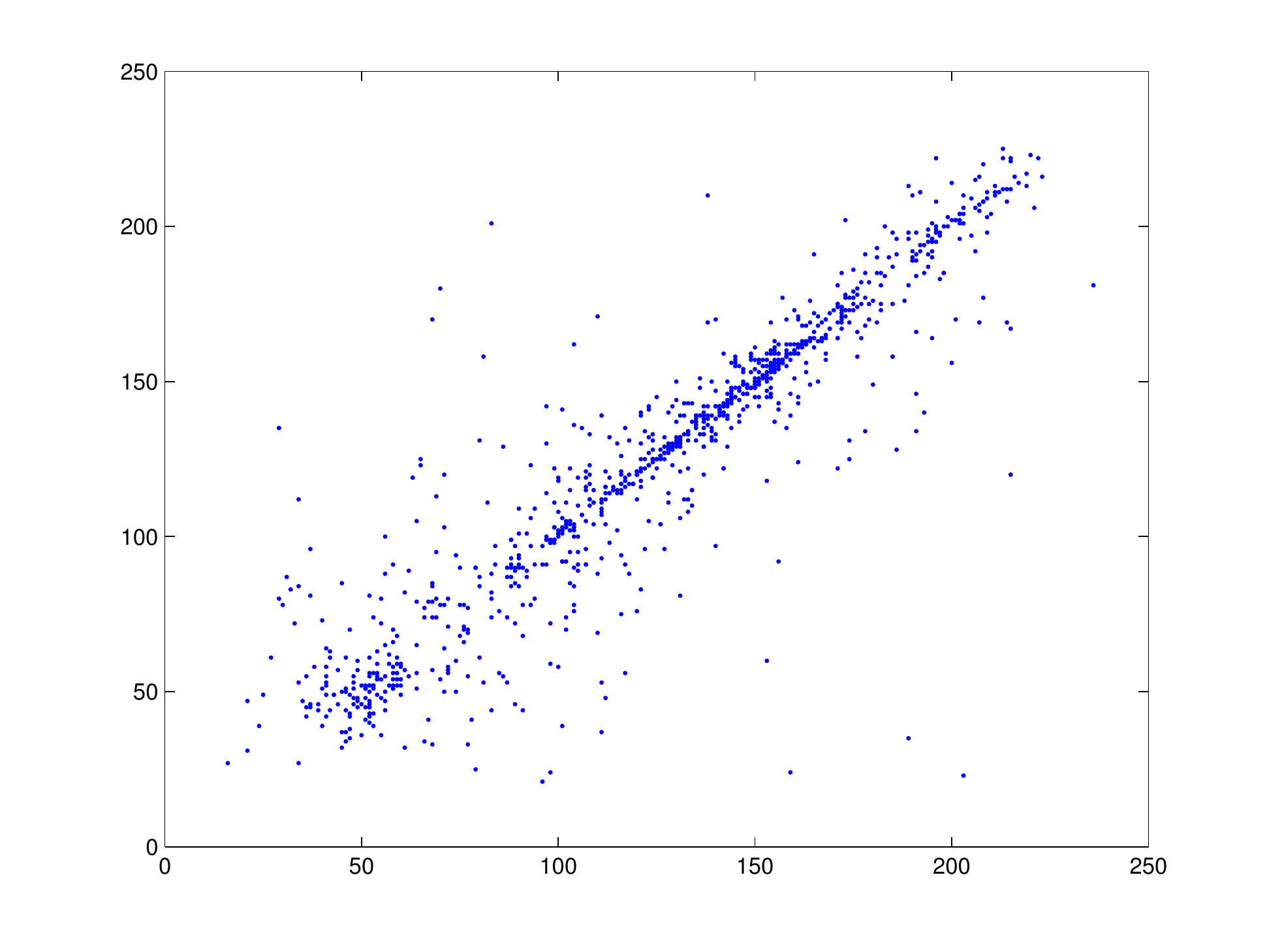}
\includegraphics[scale = 0.15]{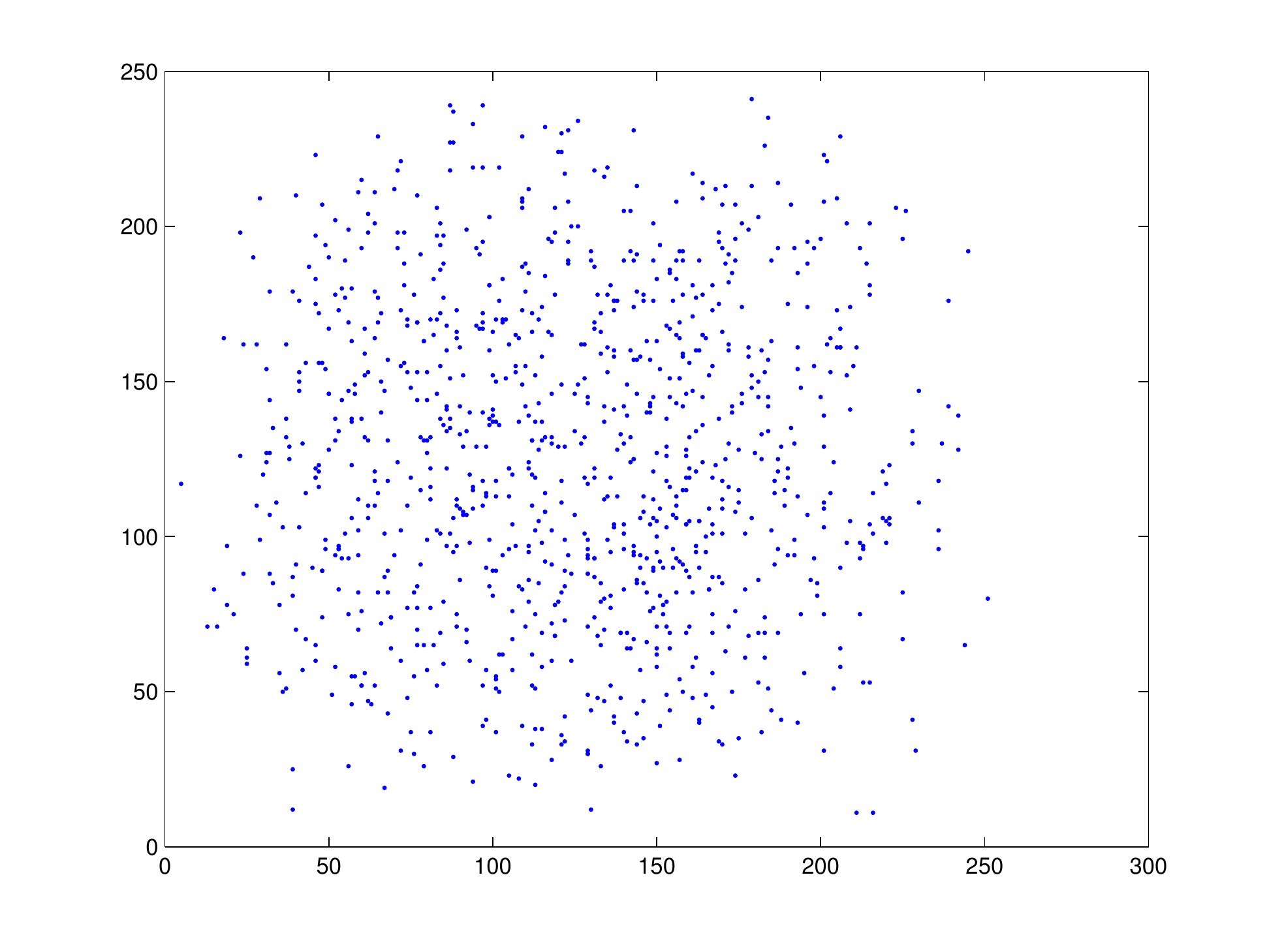}\\
\vspace{1cm}
\large ({\bf c})
\includegraphics[scale = 0.15]{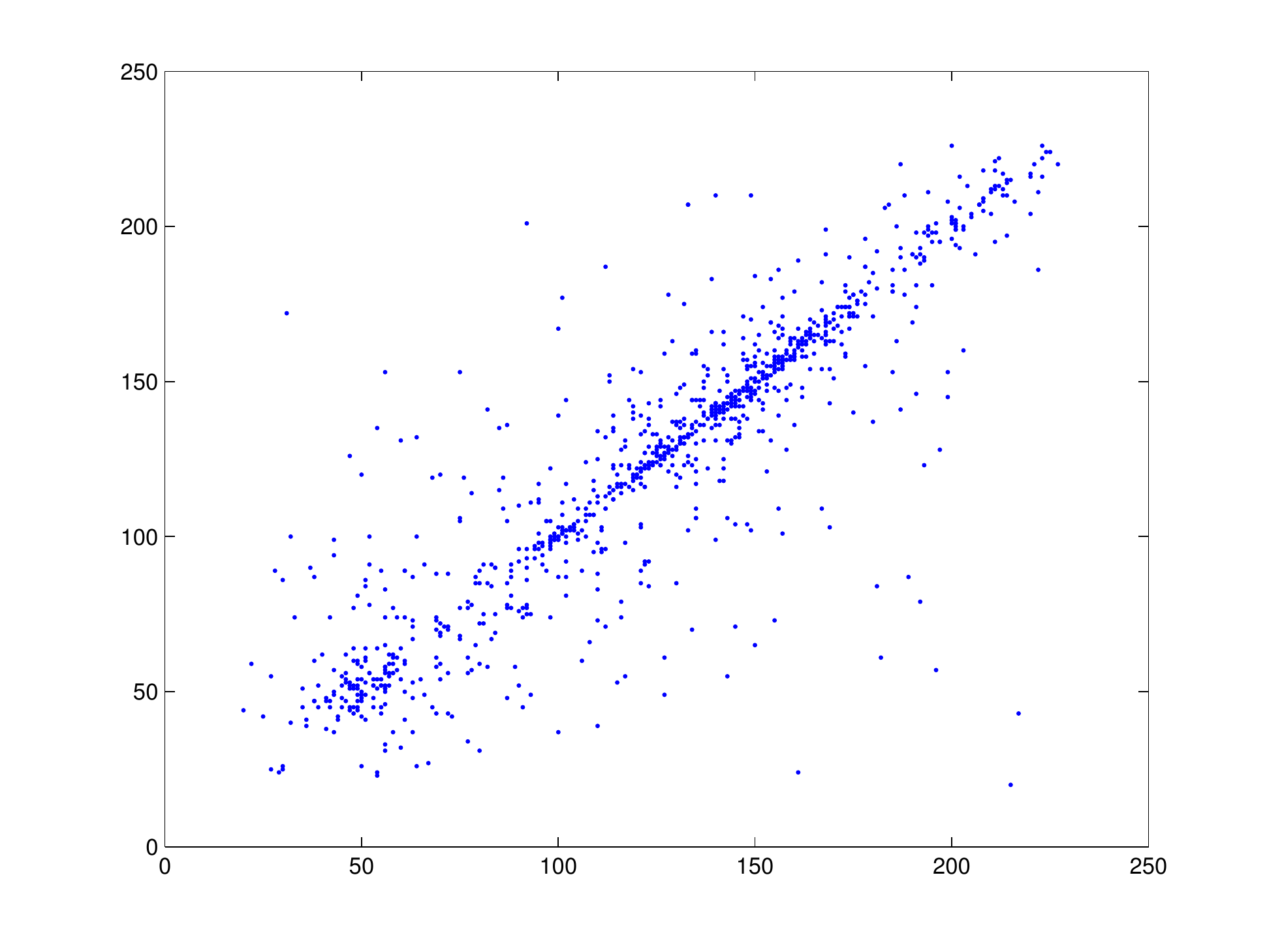}
\includegraphics[scale = 0.15]{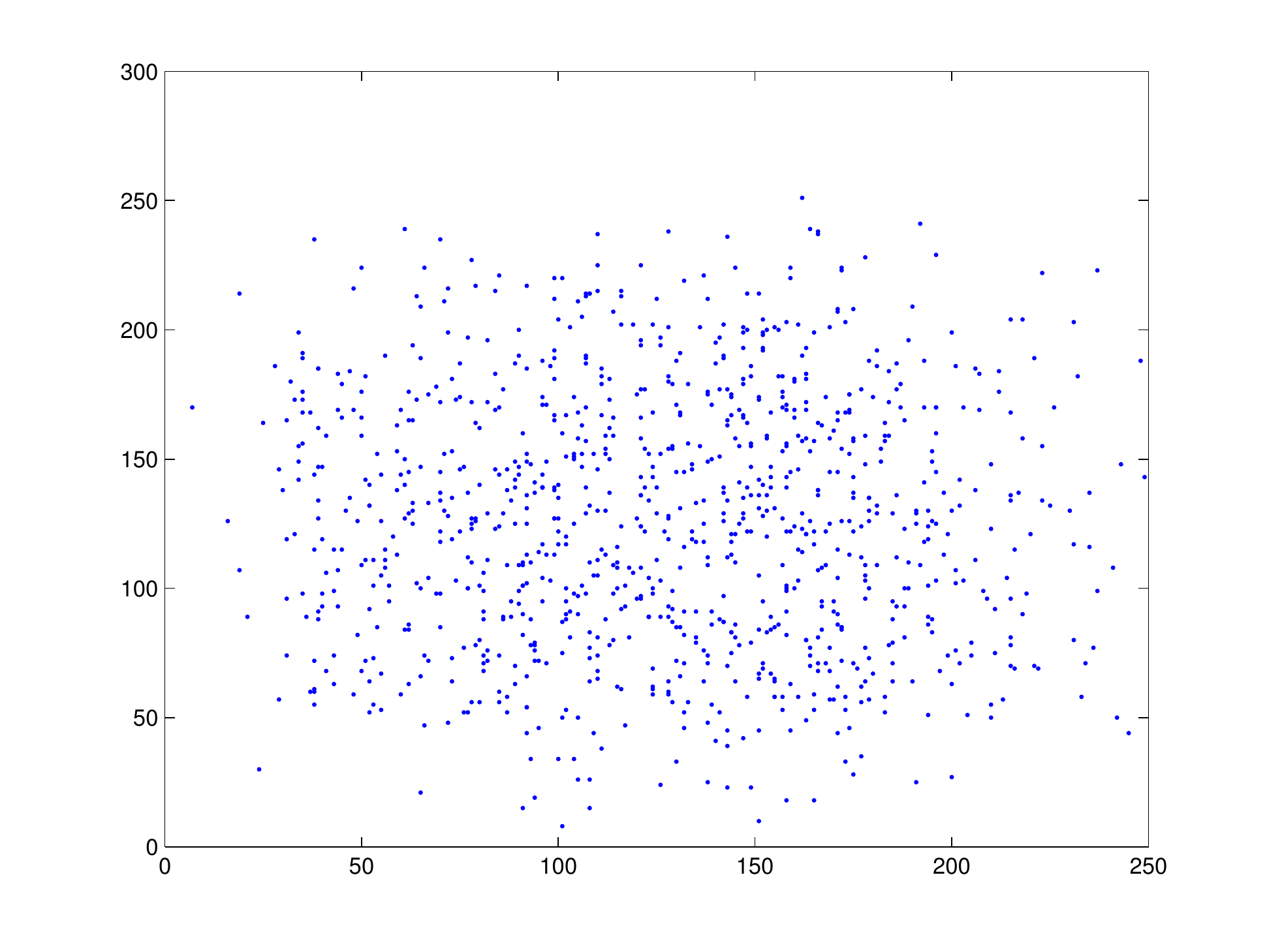}
\caption{Correlation Distribution of the pairs of adjacent pixels: {\bf(a)} Horizontal. {\bf(b)} Vertical. {\bf(c)} Diagonal }
\label{overflow}
\end{figure}

\subsubsection{Non-Parametric Test for Pixel Randomness}
Evaluation results of the proposed non-parametric test for the house images are given in figure 8. Scores of the other tests images are given in the table 2. As one can see that the perfectly encrypted and perfectly shuffled images have high scores. In contrast the original image has very low score and the partially shuffled and encrypted images have average scores. Scores of the different images are varied according to their pixel randomness. Images with high scores are indistinguishable from random patterened images. 
\begin{figure}
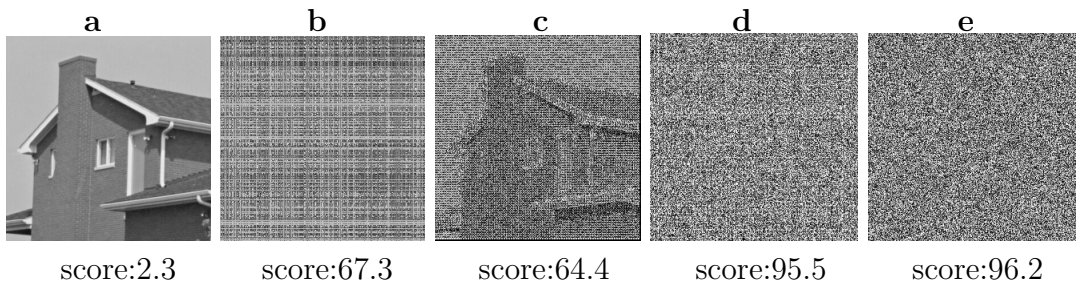

\hspace{1.3cm}{\bf a} \hspace{2.6cm}{\bf b} \hspace{2.6cm}{\bf c} \hspace{2.3cm}{\bf d} \hspace{2.6cm}{\bf e}\\
\centering
\includegraphics[scale = 0.3]{house.pdf}
\includegraphics[scale = 0.3]{house_not_perf_shuf.pdf} 
\includegraphics[scale = 0.3]{house_not_perf_enc.pdf}
\includegraphics[scale = 0.3]{house_shuf.pdf} 
\includegraphics[scale = 0.3]{house_enc.pdf} \\
\hspace{0.3cm}score:2.3
\hspace{1cm}score:67.3 
\hspace{1cm}score:64.4 
\hspace{1cm}score:95.5
\hspace{1cm}score:96.2
\caption{{\bf(a)} Original {\bf(b)} Partially Shuffled {\bf(c)} Partially Encrypted {\bf(d)} Perfectly Shuffled {\bf(e)} Perfectly Encrypted}
\end{figure}
\begin{center}
\begin{small}
\begin{tabular}{|c|c|c|c|c|c|l|}
\hline
Images &  Original  & Partially Shuffled  & Partially Encrypted   & Perfectly Shuffled  &  Perfectly Encrypted \\  \hline  
Lena Image  & 1.3 & 65.4 & 65.8  & 97.6 & 98.8\\ \hline   
Boat Image &  2.1  & 61.2 & 65.7 & 98.6 & 97.5 \\ \hline
Pepper Image &  2.6  & 66.3 & 67.4 & 97.4 & 98.5\\  \hline
\end{tabular}
{\bf Table 2:}  Scores of Test Images.
\end{small}
\end{center}
\section{Conclusion}
We have introduced a new test using non-parametric method in statistics for measuring the randomness in the image pixels. It scores the images, depending upon how far their pixels are i.i.d. The newly designed 2D CA based encryption scheme first analysed by standard evaluation methods and then the encrypted images are used to validate our newly proposed non-parametric test. Simulation results clearly show that this test can be used for evaluating the quality of any image shuffling and image encryption method.The complexity of the method is $O(NThn)$, so it can be efficiently implemented. In the proposed non-parametric test we have only used the basic run test to measure the randomness.

\end{document}